\documentclass[10pt]{article}
\usepackage{amsmath}
\usepackage{amssymb}
\usepackage{graphicx}

\def\aa{A\&A}

\begin{document}

\title{{\bf The astronomical units}}
\date{\today}
\author{N. Capitaine and B. Guinot}
\maketitle
\begin{center}
SYRTE, Observatoire de Paris, CNRS, UPMC \\ 
61, avenue de l'Observatoire, 75014 Paris, France\\
e-mail: n.capitaine@obspm.fr; guinot.bernard@wanadoo.fr \\
\end{center}

\begin{abstract}
\smallskip
The IAU-1976 System of astronomical constants includes three astronomical units (i.e.~for time, mass and length). This paper reports on the status of the astronomical unit of length~(ua) and mass ($M_{Sun}$) within the context of the recent IAU Resolutions on reference systems and the use of modern observations in the solar system. We especially look at a possible re-definition of the ua as an astronomical unit of length defined trough a fixed relation to the SI metre by a defining number. 

\end{abstract}

\vspace*{0.4cm}

\noindent {\large 1. INTRODUCTION} 
\smallskip

The IAU-1976 System of astronomical constants includes three astronomical units, namely the astronomical unit of time, the day, $D$, which is related to the SI by a defining number ($D$=86400~s), the astronomical unit of mass, i.e.~the mass of the Sun, $M_{Sun}$, and the astronomical unit of length,~ua. Questions related to the definition, numerical value and role of the astronomical units have been discussed in a number of papers, e.g.~(Capitaine \& Guinot~1995), (Guinot~1995), (Huang et al.~1995), (Standish~1995, 2004) and (Klioner~2007). The aim of this paper is to report on recent views on these topics. 

The role of the astronomical unit of time, which (as is the Julian century of 36~525 days) is to provide a unit of time of convenient size for astronomy, does not need further discussion. In the dynamics of the solar system, the unit of mass appears only through the products $GM$ (dimension L$^3$T$^{-2}$). The gravitational constant $G$ having a relative uncertainty of about 10$^{-5}$, only relative masses are obtained accurately. For these reasons we restrict our discussion to the unit of length. We recall first its historical basis and current use, then we consider a reform which might be desirable.

\vspace*{0.5cm}

\noindent {\large 2. THE SI AND THE ASTRONOMICAL UNIT OF LENGTH (UA)}
\smallskip

Physicists and astronomers have a common goal: transferring the quality of the measurement of time to that of length. This can be achieved by assigning conventional values to physical constants whose dimensions are only time and length. However an essential condition is the possibility of a realization of the unit of length with a satisfactory accuracy.

In physics, the metre of the SI has been defined in 1983 from a fixed value of the speed of light. This was made possible by the progress in the measure of optical frequencies providing wavelengths and then lengths in metre by interferometry. Of course the direct use of the definition is especially useful for distances encountered in geodesy and astronomy.

In astronomy, the unit of length (ua) is, in fact, defined by the adoption  of a conventional value of $GM_{Sun}$ [The official definition in the IAU-1976 System of astronomical constants, rather obscure, refers to the Gaussian gravitational constant, $k$ (with $k^2=G$ in astronomical units), whose value has never been changed]. The realization of the ua was based on the measurement of mean motions and on the laws of Newtonian dynamics (i.e.~Kepler's third law: $n^2a ^3=GM_{Sun}$ for a planet of negligible mass).

This conceptual difference between physicists and astronomers can be explained by the poor accuracy of evaluation of distances in SI through classical astrometric measurements: parallaxes, radial velocities, annual aberration leading to relative uncertainties of order 10$^{-4}$. As long as the value $GM_{Sun}$ remains conventional, the link between the ua and the SI metre is subject to measurement and has an uncertainty.

\newpage
\medskip

\noindent {\large 3. EVOLUTION OF MEASUREMENTS AND MODELS, CONSEQUENCES}
\smallskip

High accuracy positioning in the solar system is now based mainly on range and Doppler measurements, especially for telluric bodies (Moon, Mercury, Venus, Mars, ...) (Standish~2004, Pitjeva,~2005, Fienga et al.~2006, Folkner et al.~2008). Moreover, the space-time references have been defined in General Relativity framework in 1991, with further improvements from 2000 to present. In the adopted metrics, the $GM$s of celestial bodies remain meaningful quantities.

For the telluric bodies, for which there are precise range and Doppler measurements, coordinates are obtained in metres with a very high accuracy and $GM_{Sun}$ is obtained in SI units with a relative uncertainty of 4 $\times 10^{-9}$ (according to IERS Conventions 2003). For the planets for which observations are mainly angular measurements (Jupiter, Saturn, etc.), the relative distances are determined as with old observations and the scale in SI of the global solution is provided by the $GM_{Sun}$  value obtained from telluric bodies, as above. If very precise angular measurements for some bodies of the solar system are available, it may happen that the use of the experimental value for $GM_{Sun}$ leads to less precise absolute distance in SI than the relative distances. This is quite similar to the effect of the use of the experimental value for the ua on distances expressed in ua as it is currently defined.

\vspace*{0.4cm}

\noindent {\large 4. POSSIBLE REFORM IN THE STATUS OF THE ASTRONOMICAL UNITS}
\smallskip

Basically there are two possibilities, both valid in the Newtonian context as well as in general relativity:

- $GM_{Sun}$ conventionally fixed and the SI value of the ua estimated, which corresponds to the current official status of the ua,

- $GM_{Sun}$ estimated in SI from the observations and the ua, if needed, conventionally expressed as a fixed number of SI metres, which would correspond to a reform in the status of the ua.
\smallskip

In the latter case, the use of the ua is a simple change of unit. This ua has the same character as the metre: it is defined as a unit of proper length (Huang et al.~1995). Therefore its relation to the metre is not affected by changes in the relativistic metric (introduction of scaling factors).

\vspace*{0.4cm}

\noindent {\large 5. CONCLUSION}
\smallskip

A re-definition of the ua through a fixed relation to the SI metre by a defining number (whose value should be in agreement with the current best estimate of the ua for continuity reasons) appears desirable for modern dynamics of the solar system. This would mean:

- determining experimentally $GM_{Sun}$ which would cease to be considered as ``constant",

- limiting the role of the ua to that of a unit of length of ``convenient" size for some applications.
\smallskip

Such a change of status of the ua would

- be a great simplification for the users of the astronomical constants,

- let possible variations of the mass of the Sun, and/or $G$, appear directly,

- avoid an unnecessary deviation from the SI.

\vspace*{0.4cm}

\noindent {\large 6. REFERENCES}

{
\leftskip=2mm
\parindent=-2mm

\smallskip

Capitaine, N., Guinot, B., 1995, ``Astronomical units and constants in a relativistic framework'', {\it Highlights of Astronomy}, Vol 10, IAU, 1994, I. Appenzeller (ed), 201.

Fienga A., Manche H., Laskar J., Gastineau M., 2008, INPOP96, ``A new numerical planetary ephemeris'', \aa~477, 315.

Folkner W.M., Williams J.G., Boggs D.H., 2008, Memorandum IOM 343R-08-003, Jet Propulsion Laboratory.

Guinot B., 1995, ``Le syst\`eme international d'unit\'es (SI) et les unit\'es astronomiques'', Proc. Journ\'ees 1994 ``Syst\`emes de r\'ef\'erence spatio-temporels'', N. Capitaine (ed), 21.

Huang T.-Y., Han C.-H., Yi Z.-H., Xu B.-X., 1995, ``What is the astronomical unit of length?'', \aa~298, 629.

IERS Conventions (2003), \textit{IERS Technical Note~32}, D.D.~McCarthy and G.~Petit (eds), Frankfurt am Main: Verlag des Bundesamts f\"ur Kartographie und Geod\"asie, 2004.

Klioner S., 2007, ``Relativistic scaling of astronomical quantities and the system of astronomical units'', \aa~478, 951.

Pitjeva, E.V., 2005, ``High-Precision Ephemerides of Planets EPM and Determination of Some Astronomical Constants'', Solar System Research, Volume 39, Issue 3, pp.176-186.

Standish E.M. 1995, ``Report of the IAU WGAS Sub-Group on Numerical Standards'', {\it Highlights of Astronomy}, Vol 12, IAU 1994, Appenzeller, I. (ed), 180.

Standish E.M., 2004, ``The Astronomical Unit now'', in ``Transits of Venus, New views of the Solar System and Galaxy'', Proc. IAU Coll. 196, D.W. Kurtz (ed), 183.

}
\end{document}